\documentclass[10pt,twoside,BCOR7mm,DIV15,headinclude,footexclude,
               cleardoubleempty,idxtotoc]
{scrartcl}

\usepackage{natbib}
\usepackage[font=small,labelfont=bf]{caption}
\usepackage[english]{babel}
\usepackage{graphicx}
\usepackage{hyperref}
\usepackage{scrpage2}
\usepackage{ifthen}
\usepackage{booktabs}
\usepackage{amsmath}
\usepackage{amssymb}
\usepackage{multicol}
\usepackage{float}
\usepackage{hyperref}

\hypersetup{breaklinks=true
,colorlinks=true,linkcolor=black,urlcolor=blue
,citecolor=black}

\addto\captionsenglish{%
}

\makeatletter
\renewcommand{\@biblabel}[1]{}
\renewcommand{\@cite}[2]{%
{#1\ifthenelse{\boolean{@tempswa}}{,#2}{}}}
\makeatother
\ofoot{\thepage}
\ifoot{}

\setcapindent{0em}
\setheadsepline{1pt}

\setkomafont{pagehead}{\normalfont\sffamily}
\setkomafont{pagenumber}{\normalfont\rmfamily}

\makeatletter
\newcommand{\listofcontributions}{\@starttoc{con}}

\newcommand{\l@contribution} {\@dottedtocline{1}{1.5em}{2.3em}}
\makeatother

\newenvironment{contribution}{
\setcounter{section}{0}
\setcounter{figure}{0}
\setcounter{table}{0}
}{
\newpage
\lehead{}
\rohead{}
}

\begin{document}

\setlength{\baselineskip}{2.5ex}

\begin{contribution}

\lehead{C. Georgy et al.}

\rohead{Wolf-Rayet stars as an evolved stage of stellar life}

\begin{center}
{\LARGE \bf Wolf-Rayet stars as an evolved stage of stellar life}\\
\medskip

{\it\bf C. Georgy$^1$, S. Ekstr\"om$^2$, R. Hirschi$^1$, G. Meynet$^2$, J. Groh$^2$, \& P. Eggenberger$^2$}\\

{\it $^1$Astrophysics group, EPSAM, Keele University, Lennard-Jones Labs, Keele, ST5 5BG, UK}\\
{\it $^2$Geneva Observatory, University of Geneva, Chemin des Maillettes 51, 1290 Versoix, Switzerland}

\begin{abstract}
Wolf-Rayet (WR) stars, as they are advanced stages of the life of massive stars, provide a good test for various physical processes involved in the modelling of massive stars, such as rotation and mass loss. In this paper, we show the outputs of the latest grids of single massive stars computed with the Geneva stellar evolution code, and compare them with some observations. We present a short discussion on the shortcomings of single stars models and we also briefly discuss the impact of binarity on the WR populations.
\end{abstract}
\end{center}

\begin{multicols}{2}

\section{Introduction}

Wolf-Rayet (WR) stars are known to be the very advanced stages of the life of massive stars \citep{Conti1976a}. These stars exhibit at their surface signs of hydrogen burning products through CNO cycle (for the WN stars) or even of helium burning products \citep{Crowther2007a}. To explain the existence of such objects, elements that are produced in the centre of a massive stars have to appear at the surface. This can be achieve by two different ways: a) internal mixing inside that star, allowing for transporting chemical elements from the place where they are produced to the surface, b) mass loss, that progressively remove the external layer of the star, and can ultimately uncover deep regions of the star, where nuclear burning previously occurred.

Internal mixing can be induced by the convective movements of matter in the convective regions, or by any other kind of mixing process in the radiative regions, such as rotational mixing \citep[e.g.][]{Zahn1992a}. Mass loss can be due either by stellar winds, or by mass transfer in close binaries.

In this paper, we present preliminary results showing what are the expected WR stars population in the single massive star scenario, at different metallicities. The assumptions made in our computations and the chosen prescriptions are shortly reminded in section~\ref{georgy:section_physics}. Then, the various mass limits and number ratio of WR subtypes are presented in section~\ref{georgy:section_WR}. Finally, we briefly discuss the interesting challenges that need to be resolved to make progresses in that field.

\section{Ingredients of the stellar models}\label{georgy:section_physics}

\begin{figure*}[!t]
\begin{center}
\includegraphics[width=.28\textwidth]{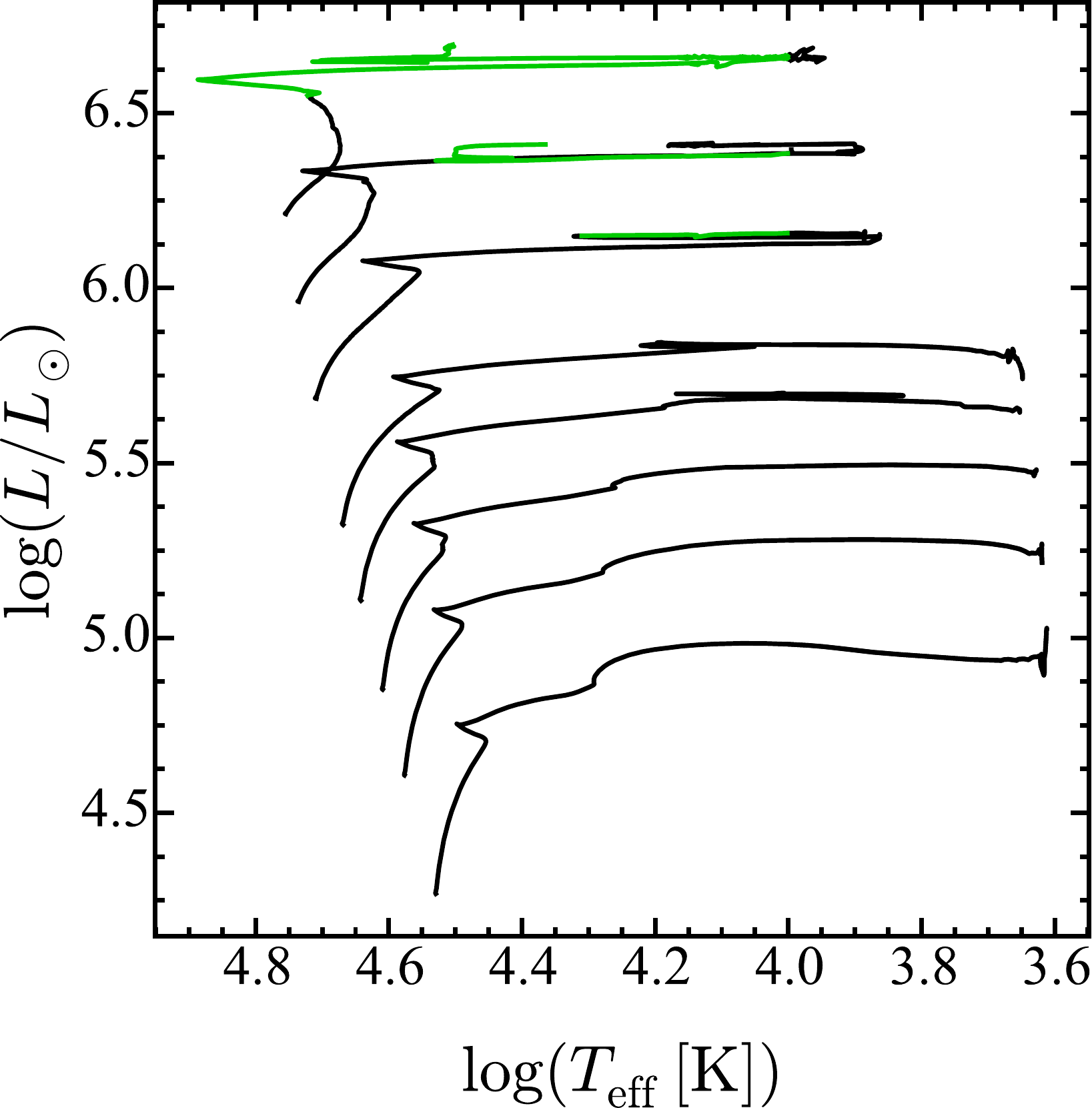}\hfill\includegraphics[width=.28\textwidth]{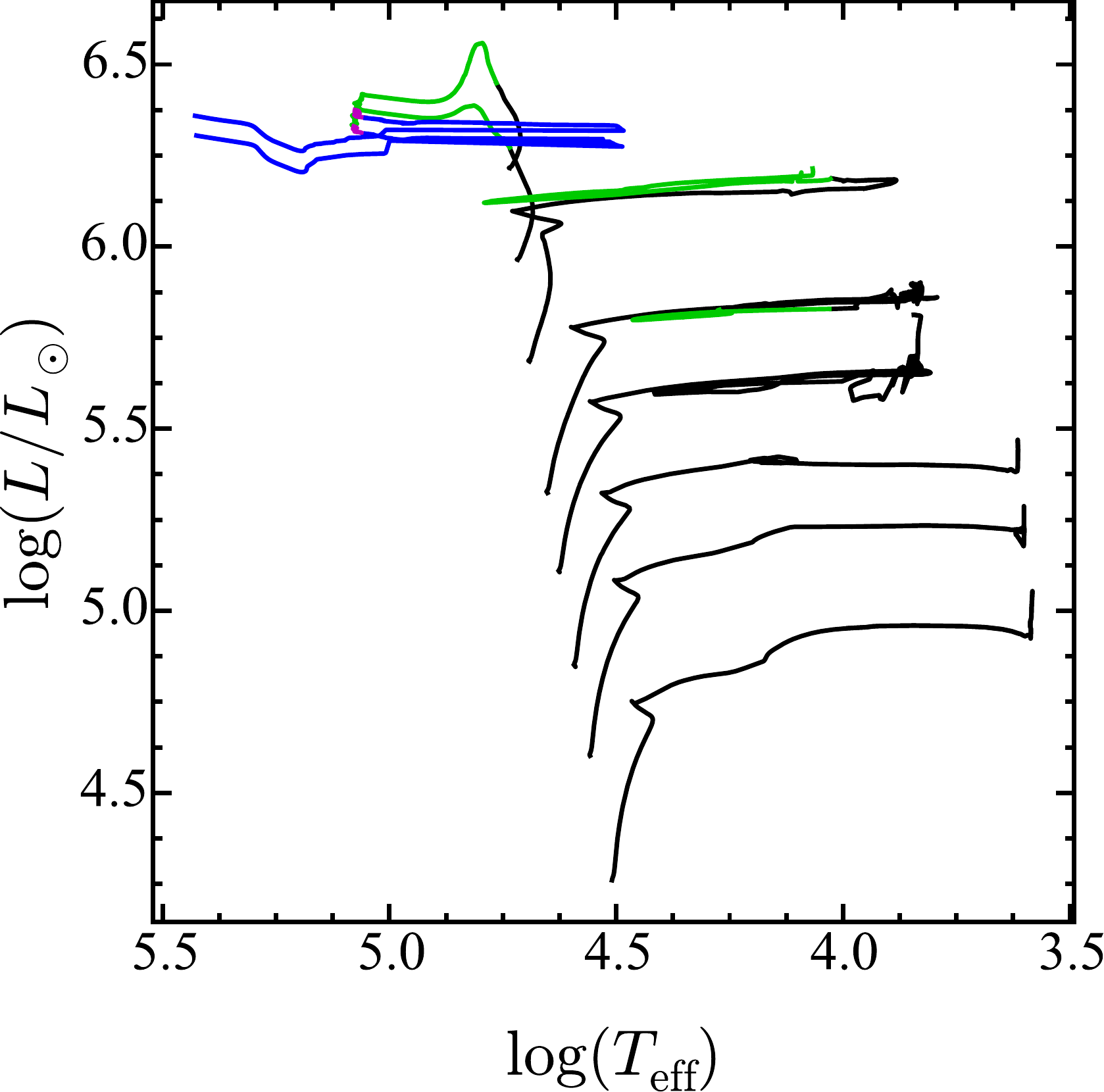}\hfill\includegraphics[width=.28\textwidth]{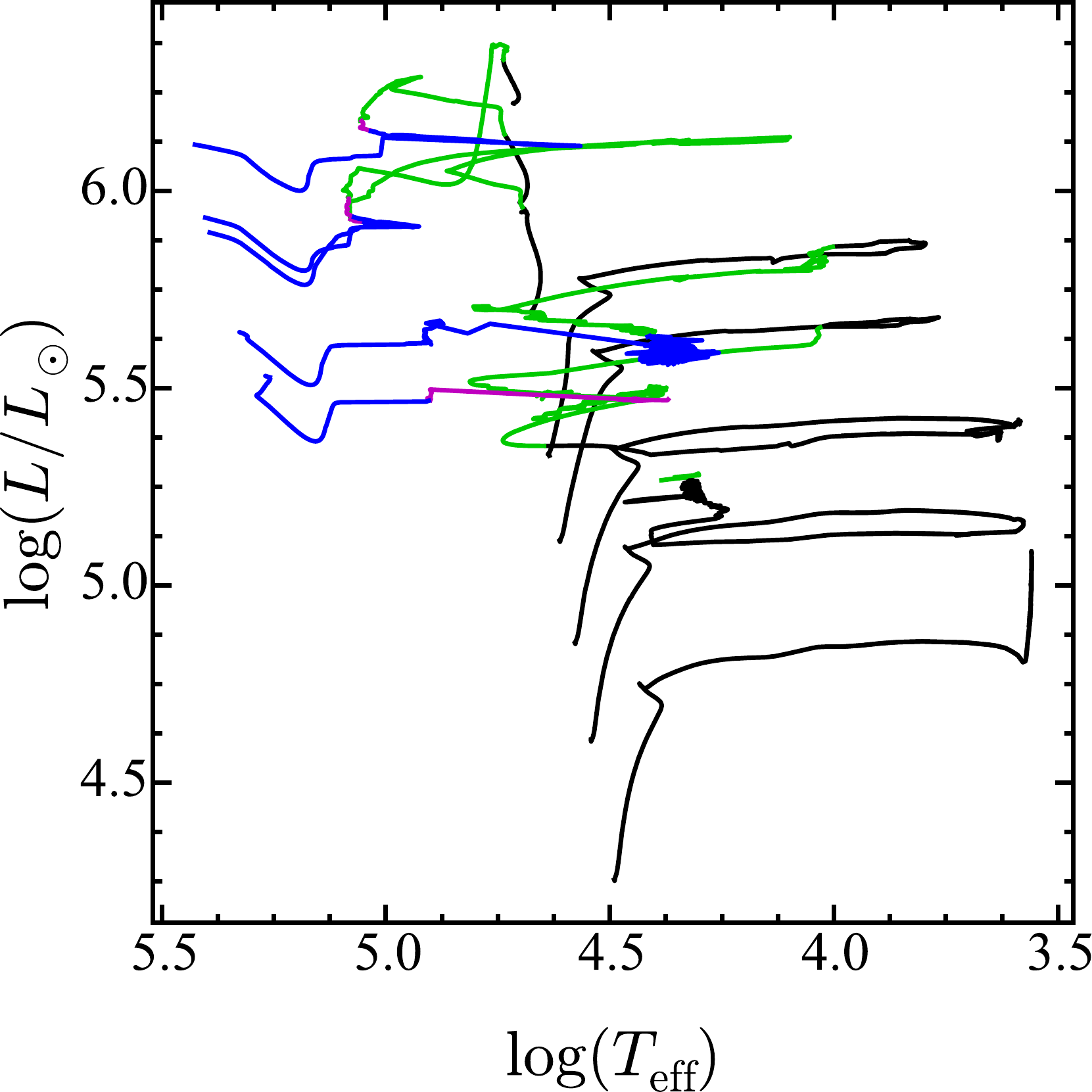}
\caption{HRD of rotating massive star models: $Z=0.002$ (SMC, \textit{left panel}), $Z=0.006$ (LMC, \textit{middle panel}), and $Z=0.014$ (solar, \textit{right pannel}). In each panel, models from $15\,M_\odot$ (bottom track) to $120\,M_\odot$ (top track) are shown. The different WR stages are indicated in colour: WNL (green), WNE (purple) and WC (blue). The initial rotation velocity is $40\%$ of the critical one. Models are taken from \citet{Georgy2012b} for $Z=0.002$, Eggenberger et al. (in prep.) for $Z=0.006$ and \citet{Ekstrom2012a} for $Z=0.014$.
\label{georgy:FigWR}}
\end{center}
\end{figure*}

The results presented in this paper were obtained by analysing the models from \citet{Ekstrom2012a}, \citet{Georgy2013b}, and Eggenberger et al. (in prep.). The computations were performed using the Geneva stellar evolution code \citep{Eggenberger2008a}. Here is a brief summary of the assumed physical ingredients and prescriptions we used.

Convective zones are determined with the Schwarzschild criterion. For hydrogen and helium burning cores, a penetrative overshoot is assumed, which extends over a fraction $d_\text{over}$ of the pressure scale height at the formal edge of the convective core. Inside the convective zones, the mixing is assumed to be instantaneous. In the inner regions, convection is assumed to be adiabatic, while in the external envelope, the thermal gradient is computed in the framework of the mixing-length theory \citep[MLT][]{Boehm-Vitense1958a}. The free parameter $d_\text{over}$ is calibrated at solar metallicity, by reproducing the width of the main sequence (MS) at around $1.7\,M_\odot$ \citep{Ekstrom2012a}. The value of the $\alpha$ parameter appearing in the MLT is calibrated on a solar model.

Rotation is included in the framework of the ``shellular'' rotation approximation \citep{Zahn1992a,Meynet1997a,Maeder1998a}. The transport of angular momentum is computed with both the advective and diffusive term \citep{Zahn1992a}, while the transport of chemical elements is a purely diffusive process \citep{Chaboyer1992a}. In this context, two diffusion coefficients are required. In our works, we use $D_\text{h}$ from \citet{Zahn1992a} and $D_\text{shear}$ from \citet{Maeder1997a}. This choice best reproduces the presence of Cepheid loops at solar metallicity. The free parameter appearing in the formulation of $D_\text{shear}$ is calibrated in order to reproduce the observed enrichment of stars in the range of $10$ to $20\,M_\odot$ at solar metallicity \citep{Ekstrom2012a}.

The mass-loss prescriptions and metallicity dependence are described in \citet{Ekstrom2012a} and \citet{Georgy2013b}. An enhanced mass-loss rate (by a factor of 3) is used for red supergiant stars (RSG) above $20\,M_\odot$ \citep[see discussion in][]{Georgy2012a,Georgy2012b}.

In this work, we use the following criteria to classify the WR stars \citep{Georgy2012b}. We consider that a star is a WR stars when its $\log(T_\text{eff})$ is above $4.$, and its surface mass fraction is smaller than $0.3$. A WR star with more than $10^{-5}$ as surface H mass fraction is a WNL star. If the surface hydrogen abundances drops below $10^{-5}$, we have a WNE star if the surface nitrogen abundance is bigger than the carbon one, and a WC star otherwise. These are obviously rough criteria, but they have the advantage that they can be computed with standard outputs from stellar evolution models. A more detailed classification would require the modelling of the emergent spectrum (taking into account the winds), and this is so far not possible during the computation runtime. Such spectrum modelling in a post-processing approach has shown that a classification based on the aspect of the spectrum leads to significant differences \citep{Groh2014a}.

\section{Wolf-Rayet stars from single star models}\label{georgy:section_WR}

Figure~\ref{georgy:FigWR} shows the Hertzsprung-Russell diagram (HRD) for our latest models of massive stars, for $15$, $20$, $25$, $32$, $40$, $60$, $85$, and $120\,M_\odot$, at three metallicities: $Z=0.002$ (\textit{left panel}), corresponding to the metallicity of the Small Magellanic Cloud, $Z=0.006$ (\textit{middle panel}), corresponding to the Large Magellanic Cloud one, and finally solar metallicity $Z=0.014$ (\textit{right panel}). The successive WR stages are highlighted with different colours: the WNL phase is shown in green, the WNE one in purple, and the WC one in blue.

A direct consequence of the weakness of the stellar winds at lower metallicity \citep{Vink2001a,Eldridge2006a} is that it becomes more and more difficult to produce WR stars when metallicity is decreasing. Moreover, at solar metallicity, our models spend a longer time in the RSG phase, allowing them to lose a lot of mass, and helping them to cross the HRD towards the WR phase at lower mass \citep[around $20\,M_\odot$, see also][]{Vanbeveren1998a}. 

The minimal masses to enter into a given WR phase are given in Table~\ref{georgy:MassLimit}. At solar and LMC metallicities, all WR types are accessible to single star models. Note however that the WC subtype occurs only for very massive stars at the LMC metallicity, while at solar $Z$, models more massive that about $27\,M_\odot$ are able to reach this stage. Moreover, these stars are found at luminosities higher than $\log(L) \sim 5.3$ at $Z_\odot$ and $\log(L) \sim 6.2$ at $Z_\text{LMC}$. At the metallicity of the SMC, the radiative stellar winds are too weak to produce stars more evolved than the WNL phase, and no stars without hydrogen at the surface are expected from our models. The minimal mass to produce a WR star progressively shift from about $20\,M_\odot$ at $Z_\odot$ to more than $50\,M_\odot$ at the SMC metallicity. Above these masses, the endpoint of the evolution of our models are thus WR stars from our simple classification scheme \citep[see however][]{Groh2013a,Groh2013b}. It is not yet clear what would be the final fate of this kind of stars in terms of supernova explosion \citep[type Ibc, failed supernova, direct collapse? See][]{Heger2003a,Dessart2011a,Groh2013c,Bersten2014a}.

\begin{table}[H]
\begin{tabular}{r c c c c }
\hline
\rule[0mm]{0mm}{3mm} &  O-star & WNL & WNE & WC \\
\cline{1-5}
Solar & $15.8\,M_\odot$ & $20.0\,M_\odot$ & $25.3\,M_\odot$ & $27.0\,M_\odot$\\
LMC & $14.2\,M_\odot$ & $32.1\,M_\odot$ & $60.8\,M_\odot$ & $63.1\,M_\odot$\\
SMC & $12.6\,M_\odot$ & $53.5\,M_\odot$ & -- & --\\
\cline{1-5}
\end{tabular}
\caption{Minimal mass to enter into a given phase. Results at $Z_\odot$ are taken from \citet{Georgy2012b}.\label{georgy:MassLimit}}
\end{table}

\begin{table*}
\begin{center}
\begin{tabular}{r c c c c c c}
\hline
\rule[0mm]{0mm}{3mm} & WR/O-stars & WNL/WR & WNE/WR & WN/WR  & WC/WR & WC/WN \\
\cline{1-7}
Solar & 0.066 &  0.687 &  0.022 &  0.709  &  0.291 & 0.409 \\
LMC & 0.016 &  0.887 &  0.005 &  0.892 & 0.108 & 0.121 \\
SMC & 0.006 &  1.000 &  0.000 &  0.000 &  0.000 & 0.000 \\
\cline{1-7}
\end{tabular}
\end{center}
\caption{Various type ratios. Values at $Z_\odot$ are taken from \citet{Georgy2012b}.\label{georgy:Ratios}}
\end{table*}

The expected ratio of WR stars to O-type stars, as well as subtype ratios, are shown in Table~\ref{georgy:Ratios}, in the constant star formation context. These numbers confirm that the number of WR stars with respect to O-type stars is decreasing with decreasing metallicity. The fraction of WC star is also expected to decrease at low metallicity, as well as the WC/WN ratio. Comparison with the observation at solar metallicity was presented in \citet{Georgy2012b}. Our results showed that in order to reproduce the observed ratio, WR stars should originate from a stellar population containing about $50\%$ of binaries, in good agreement with results from more elaborated synthetic binary population codes \citep{Vanbeveren1998a,Eldridge2008a}. Note also that our WC/WN ratio reproduce reasonably well the observed trend at different metallicities \citep{Neugent2012a}.

\section{Discussion}\label{georgy:section_conclu}

Since a few years, it became clear that massive star populations contain a significant fraction of binary stars \citep[e.g.][]{Sana2012a}. It is thus important to know what are the physical mechanisms responsible for the appearance of the WR phenomenon (or subtypes), and what is linked to binary evolution or not. Recent observations of Galactic WC stars show a significant number of such kind of stars at surprisingly low luminosities \citep[$\log(L)\sim 5.2$,][]{Sander2012a}. These stars are hard to form through single star channel, even with a strong mass-loss rate during the RSG phase \citep{Georgy2012a,Meynet2015a}, except if we assume that the strong mass loss continues \textit{after} the RSG phase \citep{Vanbeveren1998a}. On the other hand, they are routinely produced through the binary channel \citep{Eldridge2008a,Eldridge2013a}.

The relatively high number of high luminosity WNL stars with a large fraction of hydrogen on their surface \citep[e.g.][]{Hainich2014a} on the other hand points to the need of accounting for a proper treatment of the internal mixing of the star, particularly inside the radiative zones \citep{Georgy2012b}. This illustrates the need of a correct treatment of the physics of \textit{single} massive star, that definitely intervenes as well in the modelling of binary evolution.

\section*{Acknowledgements}

CG and RH acknowledge support from the European Research Council under the European Union's Seventh Framework Programme (FP/2007-2013) / ERC Grant Agreement n. 306901.

\bibliographystyle{aa} 
\bibliography{myarticle_georgy}

\begin{thebibliography}{30}
\expandafter\ifx\csname natexlab\endcsname\relax\def\natexlab#1{#1}\fi

\bibitem[{{Bersten} {et~al.}(2014){Bersten}, {Benvenuto}, {Folatelli},
  {Nomoto}, {Kuncarayakti}, {Srivastav}, {Anupama}, {Quimby}, \&
  {Sahu}}]{Bersten2014a}
{Bersten}, M.~C., {Benvenuto}, O.~G., {Folatelli}, G., {et~al.} 2014, \aj, 148,
  68

\bibitem[{{B{\"o}hm-Vitense}(1958)}]{Boehm-Vitense1958a}
{B{\"o}hm-Vitense}, E. 1958, \zap, 46, 108

\bibitem[{{Chaboyer} \& {Zahn}(1992)}]{Chaboyer1992a}
{Chaboyer}, B. \& {Zahn}, J. 1992, \aap, 253, 173

\bibitem[{{Conti}(1976)}]{Conti1976a}
{Conti}, P.~S. 1976, Memoires of the Societe Royale des Sciences de Liege, 9,
  193

\bibitem[{{Crowther}(2007)}]{Crowther2007a}
{Crowther}, P.~A. 2007, \araa, 45, 177

\bibitem[{{Dessart} {et~al.}(2011){Dessart}, {Hillier}, {Livne}, {Yoon},
  {Woosley}, {Waldman}, \& {Langer}}]{Dessart2011a}
{Dessart}, L., {Hillier}, D.~J., {Livne}, E., {et~al.} 2011, \mnras, 414, 2985

\bibitem[{{Eggenberger} {et~al.}(2008){Eggenberger}, {Meynet}, {Maeder},
  {Hirschi}, {Charbonnel}, {Talon}, \& {Ekstr{\"o}m}}]{Eggenberger2008a}
{Eggenberger}, P., {Meynet}, G., {Maeder}, A., {et~al.} 2008, \apss, 316, 43

\bibitem[{{Ekstr{\"o}m} {et~al.}(2012){Ekstr{\"o}m}, {Georgy}, {Eggenberger},
  {Meynet}, {Mowlavi}, {Wyttenbach}, {Granada}, {Decressin}, {Hirschi},
  {Frischknecht}, {Charbonnel}, \& {Maeder}}]{Ekstrom2012a}
{Ekstr{\"o}m}, S., {Georgy}, C., {Eggenberger}, P., {et~al.} 2012, \aap, 537,
  A146

\bibitem[{{Eldridge} {et~al.}(2013){Eldridge}, {Fraser}, {Smartt}, {Maund}, \&
  {Crockett}}]{Eldridge2013a}
{Eldridge}, J.~J., {Fraser}, M., {Smartt}, S.~J., {Maund}, J.~R., \&
  {Crockett}, R.~M. 2013, \mnras, 436, 774

\bibitem[{{Eldridge} {et~al.}(2008){Eldridge}, {Izzard}, \&
  {Tout}}]{Eldridge2008a}
{Eldridge}, J.~J., {Izzard}, R.~G., \& {Tout}, C.~A. 2008, \mnras, 384, 1109

\bibitem[{{Eldridge} \& {Vink}(2006)}]{Eldridge2006a}
{Eldridge}, J.~J. \& {Vink}, J.~S. 2006, \aap, 452, 295

\bibitem[{{Georgy}(2012)}]{Georgy2012a}
{Georgy}, C. 2012, \aap, 538, L8

\bibitem[{{Georgy} {et~al.}(2013){Georgy}, {Ekstr{\"o}m}, {Eggenberger},
  {Meynet}, {Haemmerl{\'e}}, {Maeder}, {Granada}, {Groh}, {Hirschi}, {Mowlavi},
  {Yusof}, {Charbonnel}, {Decressin}, \& {Barblan}}]{Georgy2013b}
{Georgy}, C., {Ekstr{\"o}m}, S., {Eggenberger}, P., {et~al.} 2013, \aap, 558,
  A103

\bibitem[{{Georgy} {et~al.}(2012){Georgy}, {Ekstr{\"o}m}, {Meynet}, {Massey},
  {Levesque}, {Hirschi}, {Eggenberger}, \& {Maeder}}]{Georgy2012b}
{Georgy}, C., {Ekstr{\"o}m}, S., {Meynet}, G., {et~al.} 2012, \aap, 542, A29

\bibitem[{{Groh} {et~al.}(2013{\natexlab{a}}){Groh}, {Georgy}, \&
  {Ekstr{\"o}m}}]{Groh2013c}
{Groh}, J.~H., {Georgy}, C., \& {Ekstr{\"o}m}, S. 2013{\natexlab{a}}, \aap,
  558, L1

\bibitem[{{Groh} {et~al.}(2013{\natexlab{b}}){Groh}, {Meynet}, \&
  {Ekstr{\"o}m}}]{Groh2013a}
{Groh}, J.~H., {Meynet}, G., \& {Ekstr{\"o}m}, S. 2013{\natexlab{b}}, \aap,
  550, L7

\bibitem[{{Groh} {et~al.}(2014){Groh}, {Meynet}, {Ekstr{\"o}m}, \&
  {Georgy}}]{Groh2014a}
{Groh}, J.~H., {Meynet}, G., {Ekstr{\"o}m}, S., \& {Georgy}, C. 2014, \aap,
  564, A30

\bibitem[{{Groh} {et~al.}(2013{\natexlab{c}}){Groh}, {Meynet}, {Georgy}, \&
  {Ekstr{\"o}m}}]{Groh2013b}
{Groh}, J.~H., {Meynet}, G., {Georgy}, C., \& {Ekstr{\"o}m}, S.
  2013{\natexlab{c}}, \aap, 558, A131

\bibitem[{{Hainich} {et~al.}(2014){Hainich}, {R{\"u}hling}, {Todt}, {Oskinova},
  {Liermann}, {Gr{\"a}fener}, {Foellmi}, {Schnurr}, \& {Hamann}}]{Hainich2014a}
{Hainich}, R., {R{\"u}hling}, U., {Todt}, H., {et~al.} 2014, \aap, 565, A27

\bibitem[{{Heger} {et~al.}(2003){Heger}, {Fryer}, {Woosley}, {Langer}, \&
  {Hartmann}}]{Heger2003a}
{Heger}, A., {Fryer}, C.~L., {Woosley}, S.~E., {Langer}, N., \& {Hartmann},
  D.~H. 2003, \apj, 591, 288

\bibitem[{{Maeder}(1997)}]{Maeder1997a}
{Maeder}, A. 1997, \aap, 321, 134

\bibitem[{{Maeder} \& {Zahn}(1998)}]{Maeder1998a}
{Maeder}, A. \& {Zahn}, J.-P. 1998, \aap, 334, 1000

\bibitem[{{Meynet} {et~al.}(2015){Meynet}, {Chomienne}, {Ekstr{\"o}m},
  {Georgy}, {Granada}, {Groh}, {Maeder}, {Eggenberger}, {Levesque}, \&
  {Massey}}]{Meynet2015a}
{Meynet}, G., {Chomienne}, V., {Ekstr{\"o}m}, S., {et~al.} 2015, \aap, 575, A60

\bibitem[{{Meynet} \& {Maeder}(1997)}]{Meynet1997a}
{Meynet}, G. \& {Maeder}, A. 1997, \aap, 321, 465

\bibitem[{{Neugent} {et~al.}(2012){Neugent}, {Massey}, \&
  {Georgy}}]{Neugent2012a}
{Neugent}, K.~F., {Massey}, P., \& {Georgy}, C. 2012, \apj, 759, 11

\bibitem[{{Sana} {et~al.}(2012){Sana}, {de Mink}, {de Koter}, {Langer},
  {Evans}, {Gieles}, {Gosset}, {Izzard}, {Le Bouquin}, \&
  {Schneider}}]{Sana2012a}
{Sana}, H., {de Mink}, S.~E., {de Koter}, A., {et~al.} 2012, Science, 337, 444

\bibitem[{{Sander} {et~al.}(2012){Sander}, {Hamann}, \& {Todt}}]{Sander2012a}
{Sander}, A., {Hamann}, W.-R., \& {Todt}, H. 2012, \aap, 540, A144

\bibitem[{{Vanbeveren} {et~al.}(1998){Vanbeveren}, {De Donder}, {van Bever},
  {van Rensbergen}, \& {De Loore}}]{Vanbeveren1998a}
{Vanbeveren}, D., {De Donder}, E., {van Bever}, J., {van Rensbergen}, W., \&
  {De Loore}, C. 1998, \na, 3, 443

\bibitem[{{Vink} {et~al.}(2001){Vink}, {de Koter}, \& {Lamers}}]{Vink2001a}
{Vink}, J.~S., {de Koter}, A., \& {Lamers}, H.~J.~G.~L.~M. 2001, \aap, 369, 574

\bibitem[{{Zahn}(1992)}]{Zahn1992a}
{Zahn}, J.-P. 1992, \aap, 265, 115

\end{thebibliography}

\end{multicols}

\end{contribution}


\end{document}